\documentclass[12pt]{iopart}

\usepackage{iopams}
\usepackage{graphicx}

\def\Rset{\mathbb R}
\def\Zset{\mathbb Z}

\newtheorem{theo}{Theorem}[section]
\newtheorem{lem}[theo]{Lemma}

\def\bmu{\boldsymbol{\mu}}

\def\esp{\mathbb E}
\def\pr{\mathbb P}

\def\rme{\mathrm{e}}

\def\erf{\mathrm{erf}}
\def\alphau{\alpha_{\textrm{\tiny u}}}
\def\alphabb{\alpha_{\textrm{\tiny bb}}}

\def\mca{\mathcal{A}}
\def\mcc{\mathcal{C}}

\def\1{\mathbf{1}}

\newcommand{\chop}{C^{\textrm{\tiny hop}}}
\newcommand{\Ibb}{C^{\textrm{\tiny bb}}} 
\newcommand{\Iloc}{C^{\textrm{\tiny loc}}} 
\newcommand{\Clocmean}{\overline{C}}

\begin{document}

\title{Topological optimization of quantum key distribution networks}

\author{R~All\'eaume$^1$, F~Roueff$^1$, E~Diamanti$^1$ and N~L\"utkenhaus$^{2,3}$}
\address{$^1$ Telecom ParisTech \& LTCI - CNRS, Paris, France}
\address{$^2$ University of Erlangen, Germany}
\address{$^3$ Institute for Quantum Computing, Waterloo, Canada}

\ead{romain.alleaume@telecom-paristech.fr}

\begin{abstract}
A Quantum Key Distribution (QKD) network is an infrastructure that allows the realization of the key distribution cryptographic primitive over long distances and at high rates with information-theoretic security. In this work, we consider QKD networks based on trusted repeaters from a topology viewpoint, and present a set of analytical models that can be used to optimize the spatial distribution of QKD devices and nodes in specific network configurations in order to guarantee a certain level of service to network users, at a minimum cost. We give details on new methods and original results regarding such cost minimization arguments applied to QKD networks. These results are likely to become of high importance when the deployment of QKD networks will be addressed by future quantum telecommunication operators. They will therefore have a strong impact on the design and requirements of the next generation of QKD devices.
\end{abstract}


\maketitle


\section{Introduction}
\label{sec:intro}

Quantum Key Distribution (QKD) is a technology that uses the properties of quantum mechanics to realize an important cryptographic primitive: key distribution~\footnote{More accurately, the primitive is that of secret key agreement using a public quantum channel and a public authenticated classical channel.}. Unlike the techniques used in traditional ``classical'' cryptography, for which the security relies on the conjectured computational hardness of certain mathematical problems, QKD security can be formally proven. Secret keys established via QKD are information-theoretically secure, which implies that any adversary trying to eavesdrop cannot obtain any information on the transmitted keys at any point in the future, even if she possesses extremely large computational resources.

The communication channels needed to perform QKD consist in an optical channel, on which well-controlled quantum states of light are exchanged, and a classical channel that is used for signaling during the quantum exchanges and for the classical post-processing phase, namely key reconciliation. Their combination forms a communication link, over which quantum key distribution allows two distant users to exchange a specific type of data, in particular secret keys. In this sense, QKD is by nature a telecommunication technology, and so \emph{QKD links} can be combined with appropriately designed nodes to form \emph{QKD networks}.

The performance of QKD links has rapidly improved in the last years. Starting from pioneering experiments in the 90s~\cite{bennett:jcrypto92}, important steps have been taken to bring QKD from the laboratory to the open field. Thanks to the continuous efforts invested in developing better QKD protocols and hardware, in parallel to the advancement of security proofs (see~\cite{gisin:rmp02,dusek:pino06,scarani:qp08} for reviews), the performance that can now be achieved, in terms of attainable communication distance, secret key generation rate and reliability, positions QKD as the first quantum information processing technology reaching a level of maturity sufficient to target deployment over real-world networks. Indeed, off-the-shelf QKD systems are now commercially available~\cite{idsqmagiq}, and the first QKD networks have recently been implemented~\cite{elliott:njp02,elliott:qp05,secoqc}.

Up till now, research in QKD has focused on building and optimizing individual systems to reach the longest possible distance and/or the highest possible secret bit rate, without taking into account the cost of such systems. However, as the perspective of deploying QKD networks becomes a reality, the question of optimal resource allocation, intrinsically linked to cost considerations, becomes relevant and important, as is the case for any telecommunication network infrastructure. It becomes therefore necessary to consider QKD from a cost perspective, and in particular study the potential trade-offs of cost and performance that can occur in this context.

Following the above arguments, we consider in this work the design of QKD networks from a topology viewpoint, and present techniques and analytical models that can be used to optimize the spatial distribution of QKD devices and QKD nodes within specific network architectures in order to guarantee a given level of service to the network users, at a minimum cost. We also study how cost minimization arguments influence the optimal working points of QKD links. We show in particular that, in the perspective of QKD networks, individual QKD links should be operated at an optimal working distance that can be significantly shorter than their maximum attainable distance.

The paper is structured as follows. In section~\ref{sec:QKDnetworks}, we define a QKD network and discuss the topology and characteristics of the network architecture that we consider in this work. We also introduce the concept of a backbone network structure. In section~\ref{sec:Optimization}, we present our calculations and results on network topological optimization based on cost arguments. In particular, we provide a comprehensive set of modeling tools and cost function calculations in specific network configurations, and discuss the effect of our results on the design of practical QKD networks. Finally, in section~\ref{sec:perspectives}, we discuss open questions and future perspectives for QKD networks.


\section{QKD networks}
\label{sec:QKDnetworks}

\noindent \emph{Definition and types of QKD networks}

\noindent Extending the range of quantum key distribution systems to very long distances, and allowing the exchange of secret keys between multiple users necessitates the development of a network infrastructure connecting multiple individual QKD links. Indeed, QKD links are inherently only adapted to point-to-point key exchange between the two endpoints of a quantum channel, while the signal-to-noise ratio decrease occurring with propagation loss ultimately limits their attainable range. It is then natural to consider QKD networks as a means to overcome these limitations.

A QKD network is an infrastructure composed of QKD links, \emph{i.e.} pairs of QKD devices linked by a quantum and a classical communication channel connecting two separate locations, or nodes. These links are then used to connect multiple distant nodes. Based on these resources and using appropriate protocols, this infrastructure can enable the unconditionally secure distribution of symmetric secret keys between any pair of legitimate users accessing the network.

QKD networks can be categorized in two general groups~\cite{salvail:jcs09}: networks that create an end-to-end quantum channel between the two users, and networks that require a transport of the key over many intermediate trusted nodes. In the first group, we find networks in which a classical optical function such as switching or multiplexing is applied at the node level on the quantum signals sent over the quantum channel. This approach allows multi-user QKD but cannot be used to extend the key distribution distance. Much more advanced members of this group are the quantum repeater based QKD networks. Quantum repeaters~\cite{briegel:prl98} can create a perfect end-to-end quantum channel by distributing entanglement between any two network users. The implementation of quantum repeaters, however, requires complex quantum operations and quantum memories, whose realization remains an experimental challenge. The same is true for the simpler version of quantum repeaters, namely quantum relays~\cite{collins:jmo05}, which on the one hand do not require a quantum memory but on the other cannot arbitrarily extend the QKD communication distance.\\

\noindent \emph{Trusted repeater QKD networks: characteristics and assumptions}

\noindent In this work, we are interested in the second group of networks, which we call \emph{trusted repeater QKD networks}. In these networks, the nodes act as trusted relays that store locally QKD-generated keys in classical memories, and then use these keys to perform long-distance key distribution between any two nodes of the network. Therefore, trusted repeater QKD networks do not require nodes equipped with quantum memories; they only require QKD devices and classical memories as well as processing units placed within secure locations, and can thus be deployed with currently available technologies. Indeed, the implementation of such networks has been the subject of several international projects~\cite{elliott:qp05,secoqc,dianati:scn08, peev:inprep09}.

As we will see in detail in the following section, the analysis of trusted repeater QKD networks from a topology viewpoint and with the goal of achieving optimization based on cost considerations involves modeling several characteristics of such a network, namely the \emph{user distribution}, the \emph{node distribution}, the \emph{call traffic}, and the \emph{traffic routing}. The user and node distributions, denoted by $\Pi$ and $M$ respectively, will be considered as Poisson stochastic point processes, and will be thus modeled using convenient stochastic geometry tools. Modeling the traffic demand is particularly subtle because of the variation with respect to time and distance that this traffic may feature in a real network. Calculations here will neglect these variations and will be performed under the assumption of a uniform call volume between any pair of users, denoted as $V$.

Finally, routing in trusted repeater QKD networks is performed according to the following general principle: First, local keys are generated over QKD links and are stored in nodes that are placed on both ends of each link. Global key distribution is then performed over a QKD path, \emph{i.e.} a one-dimensional chain of trusted relays connected by QKD links, establishing a connection between two end nodes. Secret keys are forwarded, in a hop-by-hop fashion, along these QKD paths. To ensure their secrecy, one-time pad encryption and information-theoretically secure authentication, both realized with a local QKD key, are performed. End-to-end information-theoretic security is thus obtained between the end nodes, provided that the intermediate nodes can be trusted.\\

\noindent \emph{Quantum backbone network architecture}

\noindent Introducing hierarchy into network design can be an extremely convenient architectural tool because it allows to break complex structures into smaller and more flexible ensembles. Indeed, such hierarchical levels offer an efficient way to help solve resource allocation problems arising in networks, ranging from network routing to network deployment planning. In this work, we will associate the notion of hierarchy in QKD networks with the existence of what we will call a \emph{quantum backbone network}.

In classical networks and especially the Internet, a backbone line is a larger transmission line that carries data gathered from smaller lines that interconnect with it. By analogy with this definition, the backbone QKD network is an infrastructure for key transport that gathers the traffic of secret key from many individual QKD links. QKD backbone links and nodes clearly appear as mutualized resources shared to provide service to many pairs of users. Keeping the fruitful analogy with classical networks, we will call \emph{access QKD links} the point-to-point links used to connect QKD end users to their nearest QKD backbone node.

The principle of traffic routing that we described above can be conveniently transposed in the context of backbone networks. In this case, traffic from individual users is gathered locally to backbone QKD nodes. This mutualized traffic is then routed hop-by-hop over the backbone structure. Furthermore, it is important to note that the node and user point process distributions are distinct when a backbone network is considered, which might not be the case in a network without backbone.\\

In the following, we will derive cost functions for different QKD network configurations, under the above assumptions regarding the topology and the way traffic is routed in these networks, and as a function of the characteristics of individual QKD links. We will then use the results to discuss how QKD networks should be dimensioned, the optimal working points of QKD links, as well as the interest of adopting a hierarchical architecture, materialized by the existence of a backbone, in QKD networks.


\section{Topological optimization based on cost arguments}
\label{sec:Optimization}


\subsection{QKD links: characterizing the rate versus distance}
\label{subsec:qkdlinkrate}

The main element underlying the cost optimization related to the deployment of quantum networks is the intrinsic performance of QKD links. This performance can essentially be summarized by the function $R(\ell)$, which gives the rate, in bit/s, of secret key that can be established over a QKD link of length $\ell$.

Clearly, this secret key bit rate varies from system to system and comparisons between systems are thus difficult to establish. Moreover, comparisons have to be related to the security proofs for which the secret key bit rates have been derived. Security proofs are not yet fully categorized, although important steps in this direction have been taken~\cite{scarani:qp08}.

As shown on figure~\ref{fig:RateQKDLink}, the typical curve describing the variation with distance of the logarithm of the mean rate of secret bit establishment $R(\ell)$ can be essentially separated into two parts:

\begin{figure}[!h]
\begin{center}
\includegraphics[width=11.5cm]{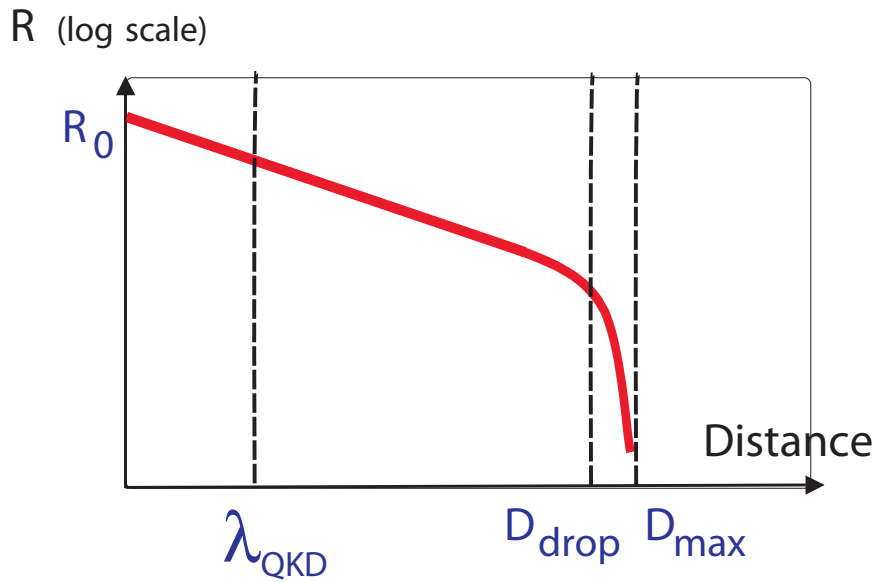}
\caption{Typical profile of the Rate versus Distance curve for a single QKD link. }
\label{fig:RateQKDLink}
\end{center}
\end{figure}

\begin{itemize}
\item A {\bf linear} part that is the region where the rate of secret key establishment varies as a given power of the propagation attenuation. Since the attenuation $\eta(\ell)$ is exponentially increasing with distance, $\log R(\ell)$ is linear in $\ell$.
\item An {\bf exponential drop-off} at longer distances, where the error rate rapidly increases due to the growing contribution of detection dark counts. In this region, the decrease of the secret key rate is multi-exponential with distance. The slope of the curve representing $\log R(\ell)$ is thus becoming increasingly steep until a maximum distance is reached.
\end{itemize}

For completeness, it is also important to mention the possibility that, for short distances, the secret bit rate could be limited by a saturation of the detection setup. This will be the case if the repetition rate at which the quantum signals are sent in the quantum channel exceeds the bandwidth of the detector. We will however not investigate this possibility any further in the remaining of this work.

The behavior of the secret bit rate function $R(\ell)$ can be described using essentially three parameters, schematically shown on figure~\ref{fig:RateQKDLink}:

\begin{enumerate}
\item The secret bit rate at zero distance, $R_0$;
\item The scaling parameter $\lambda_{\textrm{\tiny QKD}}$ in the linear region such that $R(\ell)= R_0 \, e^{-\ell/\lambda_{\textrm{\tiny QKD}}}$;
\item The distance at which the scaling of the rate becomes exponential,  which is comparable to the maximum attainable distance,  $D_{\textrm{\tiny drop}} \sim D_{\textrm{\tiny max}}$.
\end{enumerate}

$R_0$ is determined by the maximum clock rate of the QKD system. In QKD relying on photon-counting detection setups, $R_0$ is limited by the performance of the detectors, and is usually in the Mbit/s range. Clearly, the solutions allowing to improve the performance of the detectors have a direct impact on $R_0$~\cite{diamanti:pra05,yuan:apl07,hadfield:oe05,ma:ieeecl07}. For QKD systems relying on continuous variables~\cite{grosshans:nature03}, based on homodyne detection performed with fast photodiodes, the experimental bound on $R_0$ can be significantly higher, potentially in the Gbit/s range. The computational complexity of the reconciliation however currently limits $R_0$ in the Mbit/s range in the practical demonstrations performed so far~\cite{lodewyck:pra07}.

The scaling parameter $\lambda_{\textrm{\tiny QKD}}$ is essentially determined by the attenuation $\eta(\ell)$ over a quantum channel of length $\ell$, and by a coefficient $r$ that is mainly related to the security proof that can be applied to the experimental system. In the case of a typical network based on optical fibers, the attenuation $\eta(\ell)$ can be parametrized by an attenuation coefficient $\alpha$ (in dB/km) as $\eta(\ell)= 10^{- \alpha \ell/10}$ (for scaling of the attenuation in free space, see~\cite{scarani:qp08}). In the linear part of the curve shown on figure~\ref{fig:RateQKDLink}, the rate $R(\ell)$ varies as a given power $r$ of the attenuation, $R(\ell)= R_0 \, \eta(\ell)^r$. We can thus define the scaling parameter as $\lambda_{\textrm{\tiny QKD}} = 10/(\alpha \,r\,\log(10))$. For QKD performed at telecom wavelengths, with protocols optimized for long distance operation, we can take $\alpha = 0.22$~dB/km and $r=1$, which leads us to $\lambda_{\textrm{\tiny QKD}} = 19.7$~km, as the typical scaling distance for such QKD systems. This parameter is important since, as we shall see in the following, the optimal working distance of QKD links will essentially scale as $\lambda_{\textrm{\tiny QKD}}$.

Finally, the existence of a rapid drop-off of the secret key rate at distances around $D_{\textrm{\tiny drop}}$ arises when the probability to detect some signal sent in the quantum channel, $p_s$, becomes comparable to the probability to detect a dark count per detection time slot, $p_d$. This occurs around the distance $D_{\textrm{\tiny drop}}$, for which we have $p_s \simeq \exp(-D_{\textrm{\tiny drop}}/\lambda_{\textrm{\tiny QKD}}) \times \eta_{d}$, where $\eta_{d}$ represents the detector efficiency. We thus find $D_{\textrm{\tiny drop}} \simeq \lambda_{\textrm{\tiny QKD}} \, \log(\eta_d/p_d)$. In practice, when working with InGaAs single-photon avalanche photodiodes (SPADs) operating at 1550~nm, the ratio $\eta_d/p_d$ is optimized by varying the different external parameters of the detector such as the temperature, gate voltage or time slot duration. The best published performances for InGaAs SPADs \cite{zbinden:apb98,kosaka:el03} report values of the dark counts $p_d \simeq 10^{-7} \, \textrm{to} \,  10^{-6}$ for a detection efficiency $\eta_d$ around $10 \%$, which leads to $D_{\textrm{\tiny drop}} \sim D_{\textrm{\tiny max}} \sim 100-120$~km for QKD systems employing such detectors. For a similar detection efficiency, the best available superconducting single-photon detectors (SSPDs) present dark counts $p_d \simeq 10^{-8} \, \textrm{to} \,  10^{-6}$ ~\cite{korneev:jstqe07}, leading to a maximum distance that can reach 140~km.


\subsection{Toy model for QKD network cost derivation: a linear chain between two users}
\label{subsec:chain}

\emph{The linear chain as a simple asymptotic model of a quantum backbone network}

\noindent As a first example of QKD network cost derivation and optimization, we will consider what we will call the linear chain scenario. In particular, we consider two users, A and B, that want to rely on QKD to exchange secret keys in a scenario that imposes the use of several QKD links:

\begin{itemize}
\item The two QKD users are \emph{very far away}:  their distance is $L = ||AB||$  with $L \gg D_{\textrm{\tiny max}}$.
\item The two QKD users are exchanging secret bits at a \emph{very high rate}. We will call $V$ the volume of calls between the two users A and B (units of $V$: bits of secret key), and will assume $V \gg R_0$.
\end{itemize}

Because of the first condition, many intermediate nodes have to be used as trusted key relays to ensure key transport over QKD links from A to B. Because of the second condition, many QKD links have to be deployed in parallel to reach a secret key distribution rate capacity at least equal to the traffic volume.

The linear chain QKD network scenario is in a sense the simplest situation in which an infrastructure such as a quantum backbone network, described in section~\ref{sec:QKDnetworks}, is required. It therefore provides an interesting toy model for cost optimization and topological considerations.\\

\noindent \emph{Cost model: assumptions and definitions}

\noindent The generic purpose of cost optimization is to ensure a given objective in terms of service, at the minimum cost. In the case of the linear chain scenario, this objective is to be able to offer a secret bit rate of $V$~bit/s between two users A and B separated by a distance $L$, while minimizing the cost of the network infrastructure to be deployed.

In this and all subsequent models, we will consider as the total cost $\mathcal{C}$ of a QKD network, the cost of the equipment to be deployed to build the network. This can be seen as a simplifying assumption, since it is common, in network planning, to differentiate between capital and operating expenditures. We have chosen here to restrict our models to capital expenditures of QKD networks and will consider that their cost is arising from two sources:

\begin{itemize}
\item The cost of QKD link equipment to be deployed. We will denote as $C_{\textrm{\tiny QKD}}$ the unit cost per QKD link. $C_{\textrm{\tiny QKD}}$ essentially corresponds to the cost of a pair of QKD devices. Note that here we implicitly assume that the deployment of optical fibers is \emph{for free}, or more precisely that it is done independently and prior to the deployment of a QKD network.
\item The cost of node equipment, which we denote as $C_{\textrm{\tiny node}}$. $C_{\textrm{\tiny node}}$ typically corresponds to the hardware cost (for example some specific kind of routers need to be deployed inside QKD nodes), as well as the cost of the security infrastructure that is needed to make a QKD node a trusted and secure location.
\end{itemize}

\begin{figure}[!h]
\begin{center}
\includegraphics[width=13cm]{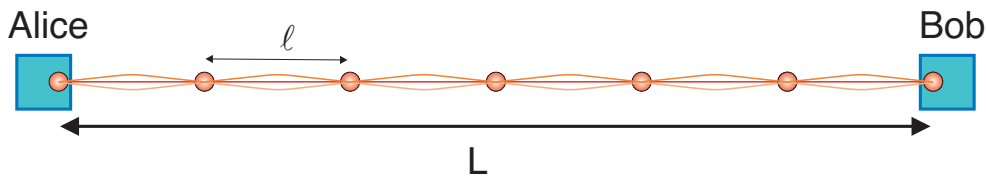}
\end{center}
\caption{The one-dimensional QKD chain linking two QKD users, Alice and Bob, over a distance $L$. Since $L$ is considered much longer than the maximum span of a QKD link, $D_{\textrm{\tiny max}}$, intermediate QKD nodes are needed to serve as trusted relays.}
\label{fig:1DQKDChain}
\end{figure}

As explained before and shown on figure~\ref{fig:1DQKDChain}, a linear chain QKD network is composed of a one-dimensional chain where adjacent QKD nodes are connected by QKD chain segments, each segment being potentially composed of multiple QKD links to ensure that a capacity equal to the traffic volume is reached.\\

\noindent \emph{Total cost of the linear chain QKD network}

\noindent For convexity reasons, discussed in more detail at the end of this section, the topology ensuring the minimum cost will correspond to place QKD nodes at regular intervals between A and B. We denote by $\ell$ the distance between two intermediate nodes, which then corresponds to the distance over which QKD links are operated within the linear chain QKD network. As we shall see, the question of cost minimization will reduce to finding the optimum value of QKD link operational distance, $\ell^{\textrm{\tiny opt}}$, for the linear chain QKD network.

There are clearly two antagonistic effects in the dependence of the total cost of the considered network on $\ell$:
\begin{itemize}
\item On the one hand, if QKD links are operated over long distances, their secret bit capacity $R(\ell)$ decreases. This will impose the deployment of more QKD links in parallel, on each chain segment linking two adjacent QKD nodes, and thus tends to increase the total cost.
\item On the other hand, it is clear that increasing the operating distance $\ell$ allows to decrease the required number of intermediate trusted relay nodes, which leads to a decreased cost.
\end{itemize}

The optimum operating distance $\ell^{\textrm{\tiny opt}}$ corresponds to the value of $\ell$ that minimizes the total cost function $\mathcal{C}$:
\begin{equation}
\mathcal{C} =  C_{\textrm{\tiny QKD}} \,  \frac{L}{\ell}  \,  \frac{V}{R(\ell)} + C_{\textrm{\tiny node}}   \frac{L}{\ell}
\label{eq:C1D}
\end{equation}

It is important to note that, in the above equation, we have made the assumption that we can neglect the effects of discretisation. This means that the length of the chain, $L$, can be considered much longer than the length of individual QKD links, $\ell$, and that the traffic volume $V$ can be considered as a continuous quantity, neglecting the discrete jumps associated to variations in the number of calls.

\vspace{0.5cm}

\noindent \emph{Cost minimization and optimum working distance of QKD links}

\noindent In the asymptotic limit of very high traffic volume $V$, the cost of nodes can be neglected in comparison with the cost of QKD devices. The expression of the total cost in equation~(\ref{eq:C1D}) then reduces to the first term, and we have the following interesting properties:
\begin{itemize}
\item The total cost is directly proportional to the product of the traffic volume $V$ and the total distance $L$.
\item Optimizing the total cost $\mathcal{C}$ is equivalent to minimizing $C(\ell)/\ell$ where $C(\ell) = C_{\textrm{\tiny QKD}}/R(\ell)$ is
 the per-bit cost of one unit of secret key rate.
\end{itemize}

Furthermore, assuming that QKD links are operated in the linear part of their characteristic (see figure~\ref{fig:RateQKDLink}), we can write $C(\ell) = \frac{C_{\textrm{\tiny QKD}}}{R_0} e^{\, \ell/\lambda_{\textrm{\tiny QKD}}}$. Then, the value of $\ell^{\textrm{\tiny opt}}$ that minimizes the quantity $C(\ell)/\ell$ can be explicitly derived as
\begin{equation}
\ell^{\textrm{\tiny opt}} = \lambda_{\textrm{\tiny QKD}} \; ,
\end{equation}
where $\lambda_{\textrm{\tiny QKD}}$ was defined in section \ref{subsec:qkdlinkrate} as the natural scaling parameter of the function $R(\ell)$.

In the general case, the second term of the cost function in equation~(\ref{eq:C1D}), corresponding to the cost of nodes, cannot be neglected. This second term does not depend on the volume of traffic $V$, and is always decreasing with $\ell$. As a consequence, the optimum operating distance that minimizes $\mathcal{C}$ will always be greater than $\lambda_{\textrm{\tiny QKD}}$, the value minimizing the first term in equation~(\ref{eq:C1D}).

Under the assumption that the optimum distance will remain in the linear part of the function $\log R(\ell)$, we can derive the following implicit relation for $\ell^{\textrm{\tiny opt}}$:

\begin{equation}
\ell^{\textrm{\tiny opt}} =  \lambda_{\textrm{\tiny QKD}} \, \Big( 1 +  \frac{C_{\textrm{\tiny node}} }{C_{\textrm{\tiny QKD}}} \,  \frac{R_0}{V}   e^{\, -\ell^{\textrm{\tiny opt}}/\lambda_{\textrm{\tiny QKD}}}   \Big)
\label{eq:LoptWithNode}
\end{equation}

The above equation allows for a quantitative discussion of the ``weight'' of the nodes in the behavior of the cost function. Indeed, we can see that the influence of the node cost is potentially important and can lead to an optimum working distance that can be significantly greater than $\lambda_{\textrm{\tiny QKD}}$ when $ \frac{C_{\textrm{\tiny node}} }{C_{\textrm{\tiny QKD}}} \,  \frac{R_0}{V} \gg 1$.\\

\noindent \emph{Existence of an optimum working distance and convexity of $C(\ell)$}

\noindent In most of the explicit derivations performed in this work, we assume a purely linear dependency of $\log R(\ell)$ on $\ell$. This assumption is convenient but remains an approximation since it does not take into account the drop-off of $R(\ell)$ occurring around $D_{\textrm{\tiny drop}}$.

It is however possible to demonstrate the existence of an optimum working distance for QKD links in a more general case, by solely relying on the assumption that the function $R(\ell)$ is log-concave, \emph{i.e.} that $\log R(\ell)$ is concave. The log-concavity of $R(\ell)$ can be checked on a simple model inspired by the secret key rate formula for the BB84 QKD protocol with perfect single photons~\cite{scarani:qp08}. In particular, in this case we have $R(p)= 1 - 2 h(p)$, where $h(p)$ is the entropy associated to a quantum bit error rate $p$, and assume that the dependence of the error rate $p$ on the distance is of the form $p = a + b / \eta(\ell) = a  + b ^{\, \ell/\lambda_{\textrm{\tiny QKD}}} $, where $a$ and $b$ are parameters linked to the detection system. In this setup, it is straightforward to verify numerically that $\log R(\ell)$ is concave for all reasonable values of $a$ and $b$.

Since $C(\ell)$, the per-unit cost of secret bit rate on a QKD link, is proportional to $1/R(\ell)$, the log-concavity of $R(\ell)$ implies the log-convexity of $C(\ell)$, which itself implies the convexity of $C(\ell)$. Finally, we can write the total cost of the linear chain QKD network as the sum of the cost of each chain segment and the cost of the node equipment, namely
$$
\mathcal{C}(\ell_0,\dots,\ell_n) = V  \,  \sum_{i=0}^{n}  C(\ell_i)  + n\, C_{\textrm{\tiny node}} \;.
$$
In the above equation, $\ell_0$ denotes the distance between A and the first node, $\ell_k$, $k=1,\dots n-1$, the distance between the $k$th node and the $k+1$th node, and $\ell_n$  the distance between the last node and B. For a convex function $C$, the minimization of $\sum_{i=0}^{n}  C(\ell_i)$ under the constraint $\sum_{i=0}^{n} \ell_i=L$, where $L$ is the distance between A and B, is obtained with $\ell_i=L/(n+1)$ for all $i$. Once we set  $\ell_i=L/(n+1)$, the cost expression in the above equation only depends on $n$, or equivalently on $\ell=L/(n+1)$. For large $L$, we can disregard the fact that $\ell$ is an integer divider of $L$ and approximate $(n+1)/n$ by 1, which then leads to equation~(\ref{eq:C1D}).


\subsection{Cost of QKD networks: towards more general models}
\label{subsec:costgeneral}

The linear chain toy model developed in section~\ref{subsec:chain} provides an interesting intuition into the behavior of the cost function. The most important result is that, in the limit of large traffic rates and/or low cost of QKD nodes, the QKD network cost optimization reduces to the minimization of $C(\ell)/\ell  \sim 1 / (R(\ell) \ell)$. This leads to the existence of an optimum working distance, $\ell^{\textrm{\tiny opt}}$, at which QKD links need to be operated in order to minimize the global cost of the network deployment.

The linear chain QKD network model is however too restrictive in many aspects: it is one-dimensional and limited to the description of a network providing service to two users. We will now consider more general models, which allow us to study the more realistic case of QKD networks spanning a two-dimensional area, and providing service to a large number of users.\\

\noindent \emph{Modeling network spatial processes with stochastic geometry}

\noindent Stochastic geometry is a very useful mathematical tool for modeling telecommunication networks. It has the advantage of being able to describe the essential spatial characteristics of a network using a small number of parameters~\cite{baccelli:ts97}. It thus allows to study some general characteristics of a given network, like the behavior of its cost function, under a restricted set of assumptions. This approach fits well with the objectives of this work, and so we have employed stochastic tools to model a QKD backbone network.

As we shall see, instead of calculating the cost of a QKD network for fixed topologies and traffic usage, we will try to understand the general behavior of the cost function by calculating the \emph{average} cost function, where the average will be taken over some probability distributions of spatial processes modeling QKD users and QKD node locations.

\begin{figure}[!h]
\begin{center}
\includegraphics[width=9cm]{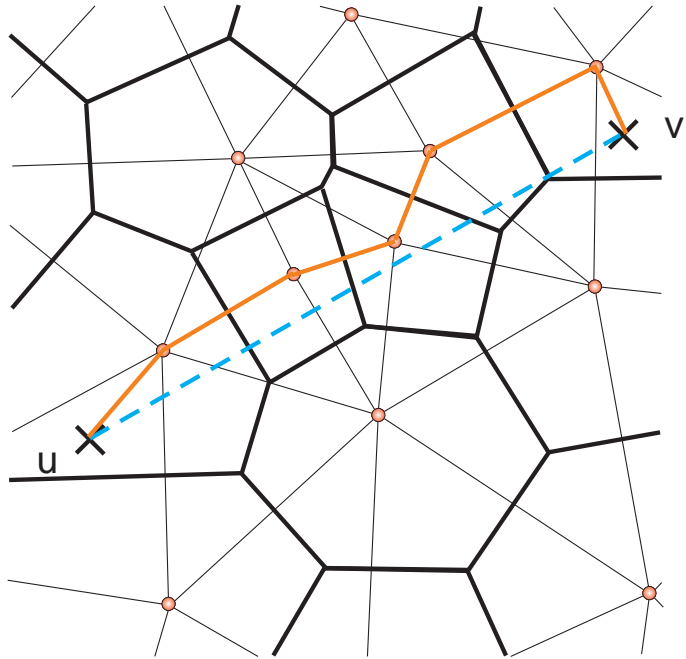}
\caption{Thick black lines: Vorono\"{\i} partition associated to a given distribution of nodes. Thin black lines: the Delaunay graph, connecting the center of neighboring Vorono\"{\i} cells. In the backbone QKD network model, backbone QKD links will indeed correspond to the Delaunay graph, while backbone nodes correspond to the nucleus of the Vorono\"{\i} cells. We have also represented on the same figure a typical end-to-end path, between two QKD users $u$ and $v$, under the Markov-path routing policy (see text in section~\ref{subsec:stochasticQBB} for details).}
\label{fig:Voronoi}
\end{center}
\end{figure}

The collection of spatial locations of the QKD nodes over the plane will be represented by a spatial point process $M=\{ X_i\}$. Then, as illustrated in figure~\ref{fig:Voronoi}, we define a corresponding partition of the plane~\footnote{More accurately, the geometrical object we consider here is a tesselation, the boundaries of which are neglected.} as the ensemble of the convex polygons $\{ D_i\}$, known as the Vorono\"{\i} cells of nucleus $\{ X_i\}$. Each Vorono\"{\i} cell $ D_i$ is constructed by taking the intersection of the half-planes bounded by the bisectors of the segment $[X_i, X_j]$ and containing $ X_i$. The system of all the cells creates the so-called Vorono\"{\i} partition. Finally, we define the Delaunay graph as the graph, whose vertices are the $\{ X_i\}$ and whose edges are formed by connecting each Vorono\"{\i} cell nucleus $\{ X_i\}$ with the nuclei of the adjacent Vorono\"{\i} cells.\\

\noindent \emph{User distribution and traffic}

\noindent In the remaining of this paper, and in contrast to the linear chain toy model developed in section~\ref{subsec:chain}, we will consider QKD networks providing secret key distribution service to a large number of users, distributed over a two-dimensional area.

The user distribution will be modeled by a Poisson stochastic point process, $\Pi=\{U_i\}$, defined over the support $D$ of size $L \times L$, while the average number of QKD users will be denoted by $\mu$. The point process $\Pi$ will also be assumed to have an intensity density $f$ satisfying  $\mu=\int f <\infty$, which means that for every set $E$ the number of users within $E$ is a Poisson random variable with mean $\int_E f$.

Finally, whenever this additional assumption will prove to be useful to perform the desired calculations, we will consider that the distribution of users is homogeneous over $D$, \emph{i.e.} that the intensity function $f$ is constant over $D$. We will denote this constant user density by $1/\alphau^2$ so that $\alphau$ corresponds to a distance (it can be shown that for large $L$, $\alphau/2$ is the average distance between the origin and the point $U_i$ closest to the origin). We will have in this case:
\begin{equation}
  \label{eq:mudef}
\mu=\int f = \left(L/\alphau\right)^2 \; .
\end{equation}

For the traffic model, we will generalize the assumption made for the linear chain QKD network model: the traffic between any pair of QKD users will be seen as an aggregate volume of calls (expressed in units of secret key exchange rate). The volume of traffic will be assumed to be the same between any pair of users, and will be denoted by $V$.\\

\noindent \emph{QKD networks with or without a hierarchical architecture}

\noindent As was discussed in section~\ref{sec:QKDnetworks}, it is interesting to study to which extent deploying a structure such as a backbone, which is synonymous to the existence of hierarchy in a network, would be advantageous in the case of QKD networks. To this end, continuing to place ourselves in the perspective of cost optimization, we will derive cost functions for QKD network models with or without a quantum backbone. The obtained results will then allow us to establish comparisons and thus discuss the interest of hierarchy in quantum networks.


\subsection{Cost function for a two-dimensional network without backbone: the generalized QKD chain model}
\label{subsec:2Dchain}

A direct way to generalize the two-user one-dimensional chain model presented in section~\ref{subsec:chain} is simply to assume that a chain of QKD links and intermediate nodes will be deployed between each pair of users $u$ and $v$ within the QKD network. Each chain will therefore be dimensioned in order to accommodate a volume $V$ of calls. The routing of calls is trivial on such a network. The distance between the intermediate nodes on a chain will be denoted by $\ell$, as in section~\ref{subsec:chain}.

Here as well, we neglect the effects of discretisation, \emph{i.e.} the length of the chains, $||u-v||$, will be considered much longer than the length of individual QKD links, $\ell$, and the traffic volume $V$ will be considered a continuous quantity. Under these assumptions, we know that the cost associated with a pair of users located respectively at positions $u$ and $v$ and exchanging a volume $V$ of calls is (see equation~(\ref{eq:C1D}))
\begin{equation}
\mcc^{\textrm{\tiny pair}}(u,v) = V \, ||u-v|| \, C(\ell)/\ell \,+ \,  (||u-v||/\ell) C_{\textrm{\tiny node}}
\label{eq:Cchain}
\end{equation}

Recall that the distribution of users is described by a Poisson point process $\Pi=\{U_i\}$. Then, we can calculate the average total cost of the QKD network, $\mcc$, by summing up the costs $\mcc^{\textrm{\tiny pair}}(U_k,U_l)$ associated with the QKD chains deployed between each pair of users over $k\neq l$ and then average this sum over the stochastic user point process $\Pi$:
\begin{eqnarray}
\label{eq:Cchaintotal}
\mcc & = \esp \left[\sum_{k \neq l} \mcc^{\textrm{\tiny pair}}(U_k, U_l)\right] \nonumber \\
& = \esp \left[\sum_{k \neq l} V \, ||U_k-U_l|| \, C(\ell)/\ell \,+ \,  ||U_k-U_l|| C_{\textrm{\tiny node}}\right] \nonumber \\
& = ( V \, C(\ell)/\ell  \, +  C_{\textrm{\tiny node}}/\ell ) \,\delta \;,
\end{eqnarray}
where $\delta$ is the average sum of distances over all pairs of two different users, namely
\begin{equation}
 \label{eq:deltaDef}
 \delta=\esp \left[\sum_{k \neq l}  ||U_k-U_l||\right] \;.
\end{equation}
For a homogeneous Poisson point process $\Pi$ with spatial density of users $\alphau^{-2}$ over a square domain $D$ of size $L\times L$, it is possible to perform the exact integral calculation of $\delta$, yielding
\begin{equation}
 \label{eq:deltaVal}
 \delta=\gamma \,L^5/\alphau^{4}\quad\textrm{with}\quad\gamma = \frac{1}{3} \log ( 1+ \sqrt2) +  \frac{2 +  \sqrt2}{15} \simeq 0.5214\;.
\end{equation}


\subsection{Cost function for a two-dimensional QKD network with backbone}
\label{subsec:cost2Dbackbone}

The backbone architectures we will consider in this work are \emph{topological}: for a given distribution of QKD nodes, which will be either deterministic (section~\ref{subsec:square}) or stochastic (section~\ref{subsec:stochasticQBB}), the backbone cells and backbone links will strictly coincide with the Vorono\"{\i} cells and the edges of the corresponding Delaunay graph defined above, respectively.\\

\noindent \emph{Routing traffic over a QKD backbone network}

\noindent The backbone hierarchical structure provides a convenient way to solve the routing problem that we have adopted in our cost calculations. For a given origin-destination pair of users (A,B) wishing to exchange a volume of calls $V_{AB}$, the traffic is routed in the following way:
\begin{itemize}
\item The traffic goes from A to its nearest QKD backbone node $X_A$ (center of the backbone cell containing A), through a single QKD link (an access link).
\item The traffic is routed through the {\bf optimal (less costly) path} over the backbone QKD network from $X_A$ to $X_B$ (QKD node closer to B).
\item The traffic goes from $X_B$ to B.
\end{itemize}

The routing rule defined above can be characterized as \emph{geographical}, in the sense that it is driven by distance considerations. However, determining the optimal path in a given backbone network of arbitrary topology may not be a tractable problem. Even in standard networks, where the optimal path is the shortest one, an analytic computation of the average length/cost is not always possible. In the context of backbone nodes distributed as a Poisson point process, an alternative suboptimal routing policy, the so called \emph{Markov path}, has been proposed, and leads to analytic computation of the average path length. In QKD networks, the cost is a non-linear function of the length and some adjustments are required. We consider two different geometries for the backbone:
\begin{enumerate}
\item A square backbone QKD network (section~\ref{subsec:square}), \emph{i.e.} a regular structure where nodes and links form a regular graph of degree 4. In this case finding the length of the shortest path between two nodes is trivial: backbone nodes $X_A$, $X_B$ can be designated by cartesian coordinates $(x_A, y_A)$, $(x_B, y_B)$ and the shortest path length is simply $|x_A - x_B| + |y_A - y_B| $. Moreover, cost calculations are simplified using the fact that the links between two neighbor nodes of the backbone all have the same length.
\item A stochastic backbone network (section~\ref{subsec:stochasticQBB}), where backbone nodes are distributed following a random point process
    and backbone cells are the corresponding Vorono\"{\i} partition. For this stochastic backbone, we have used a routing technique called \emph{Markov-path routing} for which, as previously established by Tchoumatchenko \emph{et al.}~\cite{tchoumatchenko:phd99,baccelli:aap00}, the average length of routes can be calculated. In the following, we will adapt these calculations to our cost function $C(\ell)$.\\
\end{enumerate}

\noindent \emph{Generic derivation of the cost function for QKD backbone networks}

\noindent For a QKD network with a backbone structure, we define $M=\{X_i\}$ as the point process of the network node distribution, and $\Pi=\{U_i\}$ as the point process of the network user distribution, with intensity density $f$. Each node $X_i$ is connected to some nodes in its neighborhood and to the clients belonging to the associated cell $D_i$. In the following, we will assume that $M$ is statistically independent of $\Pi$, and that the cells $D_i$ are the Vorono\"i cells associated to $M$, that is

\begin{equation}
 \label{eq:voronoiCell}
 D_i= \left\{x\;:\;\|x-X_i\|\leq \inf_{j\neq i}\|x-X_j\|\right\}\;.
\end{equation}

In the case of the QKD backbone network, our routing policy allows to calculate $C^{\textrm{\tiny pair}}(u,v;M)$, the QKD equipment cost associated with sending one unit of call between users $u$ and $v$, over a network whose backbone nodes are described by the point process $M$:
\begin{equation*}
C^{\textrm{\tiny pair}}(u,v;M) = \left\{ \begin{array}{ll}
            C(\|u-X_i\|)+C(\|v-X_i\|) \\
            \;\;\;\;\;\;\;\;\textrm{ if } u,v\in D_i \\
            C(\|u-X_i\|)+C(\|v-X_j\|)+ C^{\textrm{\tiny hop}}(i,j;M) \\
            \;\;\;\;\;\;\;\;\textrm{ if } u\in D_i\textrm{ and }v\in D_j\textrm{ with } i\neq j \;,
            \end{array} \right.
\end{equation*}
where $C(\ell)$ is the cost spent to send a secret bit on a QKD link over a distance $\ell$ and $\chop(i,j;M)$ is the cost to send a secret bit between the nodes $X_i$ and $X_j$ of the backbone for the given routing policy.

Given that the volume between each pair of users is $V$, the average total cost $\mcc$ of the QKD network then reads
\begin{equation*}
\mcc = \mcc^{\textrm{\tiny QKD}} + \mcc^{\textrm{\tiny node}} = V \times \esp \left[\sum_{k\neq l}   C^{\textrm{\tiny pair}}(U_k,U_l;M)\right]  + C_{\textrm{\tiny node}}\,N^2  \;,
\end{equation*}
where $N^2$ is the average number of nodes of the backbone deployed in the domain $D$ of size $L\times L$. Here $\esp$ denotes the average cost over the spatial distributions of users and backbone nodes, that is over the realizations of $\Pi$ and $M$. Since $M$ and $\Pi$ are supposed independently distributed, we may compute this average successively with respect to $M$ and $\Pi$. The total cost, averaged only over $\Pi$, can be decomposed as follows:
\begin{eqnarray*}
\fl \esp \left[\sum_{k\neq l}   C^{\textrm{\tiny pair}}(U_k,U_l;M) \right] & = \int  C^{\textrm{\tiny pair}}(u,v;M) \, f(u)\,f(v) \,du\,dv \\
& = \sum_k \int_{D_k\times D_k} \left\{C(\|u-X_k\|)+C(\|v-X_k\|)\right\}\, f(u)\,f(v) \,du\,dv \\
& \hspace{0.1cm} +  \sum_{k\neq l}\int_{D_k\times D_l} \left\{C(\|u-X_k\|)+C(\|v-X_l\|)+ \chop(k,l;M)\right\}\,f(u)\,f(v) \,du\,dv \\
 & = \sum_k\sum_l \int_{D_k\times D_l} \left\{C(\|u-X_k\|)+C(\|v-X_l\|)\right\}\, f(u)\,f(v) \,du\,dv \\
& \hspace{0.1cm} +  \sum_{k\neq l}\int_{D_k\times D_l} \chop(k,l;M)\,f(u)\,f(v) \,du\,dv \\
\end{eqnarray*}
As we can see from the last expression, the total cost $\mcc$ can be separated in three terms:
\begin{equation}
\label{eq:totalCost}
\mcc =: \Iloc +\Ibb\ + \mcc^{\textrm{\tiny node}}\;,
\end{equation}
where $\Iloc$ takes into account all connections from one client to the closest backbone node, $\Ibb$ all connections from one backbone node to another, and $\mcc^{\textrm{\tiny node}}$ is the cost of node equipment. The explicit models that we will study will allow us to compare the behavior of these different terms and thus to understand how QKD network backbone topologies can be optimized.


\subsection{Cost calculations for two explicit quantum backbone models}
\label{subsec:costcalc}


\subsubsection{Cost of the square backbone QKD network}
\label{subsec:square}

\paragraph{Network model:} We consider, as a first simple example, the case of a QKD backbone network that has a perfectly regular topology, and for which the shortest path length between two backbone nodes is easily determined.

\begin{figure}[!h]
\begin{center}
\includegraphics[ width= 11 cm]{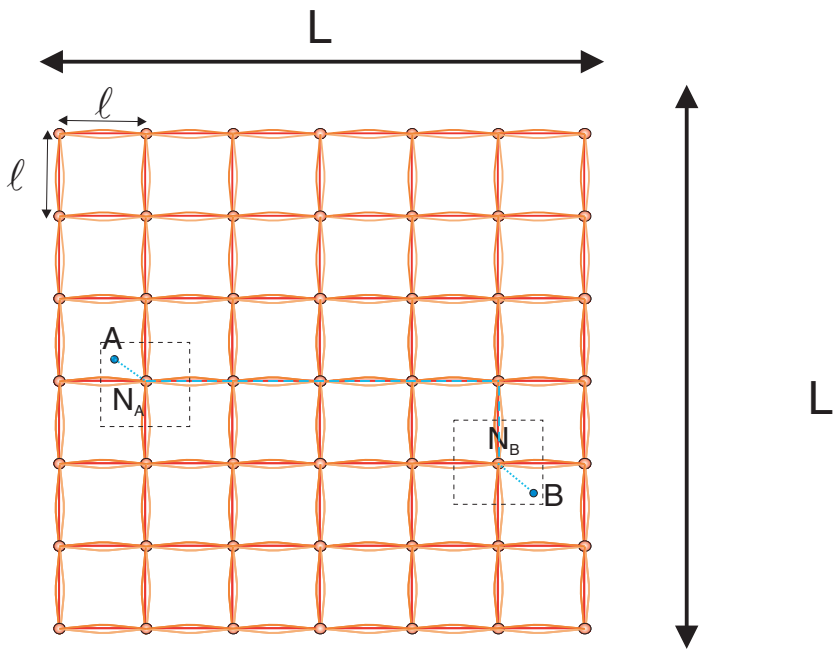}
\caption{Structure of a two-dimensional regular square backbone network: a regular array of cells of dimension $\alphabb$ spans
 a region of size $L\times L$. The user distribution is described by a random point process. In each cell, a central node collects all the local traffic. Every user in the cell is thus connected via a QKD link to the central node of its cell. On top of this array of cells, a backbone network connects first-neighbor QKD nodes with a QKD trunk. Traffic on the backbone network is routed trough the shortest path. The dotted blue line describes the path followed for the communication between two users A and B (see text for more details).}
\label{fig:SquareBB}
\end{center}
\end{figure}

The architecture we consider is the following: users are distributed as previously over a large area $D$ of size $L\times L$ and the backbone QKD network is a regular graph of degree 4, \emph{i.e.} the backbone QKD nodes and links constitute a square network. The structure of the square backbone QKD network and the way a call is routed is summarized on figure~\ref{fig:SquareBB}. The free parameter with respect to which we will perform the cost optimization is the size of backbone cells $\alphabb$. We will also make the assumption that the user density function $f$ is uniform over $D$.

\paragraph{Computation of $\Ibb $ for the square network:} We set $X_k=k\alphabb$ and $D_k=X_k+\alphabb[-1/2,1/2]^2$ with $k\in\Zset^2$ and, for all $k\neq l$,
\begin{equation*}
\chop(k,l;M) = \|k-l\|_1\, C(\alphabb) \; .
\end{equation*}
Here, $\|k-l\|_1$ corresponds to the number of hops between $X_k$ and $X_l$ and $C(\alphabb)$ to the per bit cost of one hop.

Calling $\mu_i$ the average number of QKD users in a backbone cell $i$, we have:
\begin{equation}
\label{eq:Ibb}
\Ibb  = V \sum_{k\neq l}\mu_k\mu_l \, \chop(k,l;M)
\end{equation}

Hence,
\begin{equation*}
\Ibb  =  V C(\alphabb) \, \bmu^T\Gamma\bmu  \;,
\end{equation*}
where $\bmu$ is the column vector with entries $\mu_k$, $k\in\Zset^2$, and $\Gamma$ is the Toeplitz array indexed on $\Zset^2$ with entries $\Gamma_{k,l}=\|k-l\|_1$.

Since the density of users $f$ is constant and equal to $\sigma$ on its support $D$, where
$D:=\bigcup_{k\in\{0,\dots,N-1\}^2} D_k$,  $\mu_k$ is the same for all cells $D_k$: $\mu_k= \mu/ N^2$, with $N^2$ denoting
the total number of backbone cells, and $\mu=(L/\alphau)^2$ the mean number of users over $D$ (see
equation~(\ref{eq:mudef})). Hence, we find
\begin{equation*}
\Ibb= V C(\alphabb)\, \mu^2/N^{4}\,\sum_{k,l\in\{0,\dots,N-1\}^2}\|k-l\|_1 \; .
\end{equation*}
Now, we compute
\begin{eqnarray*}
\sum_{k,l\in\{0,\dots,N-1\}^2}\|k-l\|_1 &
= & \sum_{k_1,l_1=0}^{N-1}\sum_{k_2,l_2=0}^{N-1}\sum_{i=1}^2 |k_i-l_i| \\
& = & 2 \sum_{k_1,l_1=0}^{N-1}\sum_{k_2,l_2=0}^{N-1} |k_1-l_1|  = 2 \,N^{2}\,\sum_{k,l=0}^{N-1} |k-l|\\
& = & 4 \,N^{2}\,\sum_{k=0}^{N-1} \sum_{l<k} |k-l|  = 4 \,N^{2}\,\sum_{k=0}^{N-1} \sum_{l<k} |k-l|\\
& \sim & \frac{2}3 \,N^{5}\,
\end{eqnarray*}
where the asymptotic equivalence holds as $N\to\infty$.
Using $N\sim L/\alphabb$ and equation~(\ref{eq:mudef}), we obtain, as $N\to\infty$,
\begin{equation}
\label{eq:CbbSquare}
\Ibb\sim V \, \frac{\mu^2}{N^4} C(\alphabb) \, \frac{2}3 \,N^{5} = \frac 23 \, \frac{C(\alphabb\alphau^4)}\alphabb \, L^5 \,V   = \frac 23 \, \frac{C(\alphabb)}\alphabb \, \mu^2 V \,L  \;.
\end{equation}
In the latter expression, we have four multiplicative terms:
\begin{enumerate}
\item $2/3$, a constant depending only on the dimension and the geometry of the backbone network (for a cube of dimension $d$, we could generalize our calculation and would find $d/3$);
\item $C(\alphabb)/\alphabb$, a cost function depending only on the distance $\alphabb$ between the nodes of the backbone;
\item $\mu^2 \, V$, the square of the mean number of users times the volume of call per pair of users, \emph{i.e.} in our communication model, the total volume of the communications over which the total cost is computed;
\item $L$, the size of the support of $f$, that is of the domain where the users lie.
\end{enumerate}

To understand better the derived expression for $\Ibb$, it is interesting to compare it with $ \Iloc $ and $\mcc^{\textrm{\tiny node}}$. Indeed, we can show that $\Iloc \simeq \mu^2 \,\Clocmean$, where $\Clocmean$ stands for the per-bit cost function $C$ averaged over one cell. In the case of the square network with $\alphabb\times\alphabb$ square cells, these cells are contained between two circles of radius $\alphabb/2$ and $ \alphabb \, \sqrt{2}/2 < \alphabb$. Since $C$ is an increasing function of distance we have $\Clocmean < C(\alphabb)$, and we can thus derive the important following property: {\bf In the limit of large networks, \emph{i.e.} for $L \gg \alphabb$, the backbone cost is dominant over the local cost.} We will see in the following section that this property is preserved for a backbone with randomly positioned nodes and an appropriate routing policy. Furthermore, we will see that for large $L$, the backbone node equipment cost $\mcc^{\textrm{\tiny node}}$ is negligible. Therefore, to optimize the cost~(equation~\ref{eq:totalCost}), we only need to minimize $\Ibb$. Assuming a square regular backbone, this means choosing $\alphabb$ so as to minimize $C(\alphabb)/\alphabb$, exactly as in the case of the linear chain QKD network model of section~\ref{subsec:chain}.

Hence, if we take $C(\ell) = \frac{C_{\textrm{\tiny QKD}}}{R_0} e^{\, \ell/\lambda_{\textrm{\tiny QKD}}}$, the cost is minimized for
\begin{equation}
 \label{eq:alphaOptSquareBB}
\alphabb^{\textrm{\tiny opt}}  = \lambda_{\textrm{\tiny QKD}}\;.
\end{equation}


\subsubsection{Cost calculation for a stochastic QBB with Markov-path routing}
\label{subsec:stochasticQBB}

\paragraph{} We now compute $\Iloc$ and $\Ibb$ in the case where the routing policy is the so called Markov path, as proposed in~\cite{baccelli:aap00}, where some general formulae are given for computing average costs in a general framework (see also \cite{tchoumatchenko:phd99}). The routing policy is defined as follows. First, all pairs of nodes whose cells share a common edge are connected. The corresponding graph is a Delaunay graph. Next, given two users A and B with respective positions $u$ and $v$, we define a finite sequence of the nodes $X_{k_0},X_{k_1},\dots,X_{k_n}$ in the successive cells encountered when drawing a line from $u$ to $v$. This routing policy is illustrated on figure~\ref{fig:Voronoi}.

By definition, $X_{k_0}$ and $X_{k_n}$ are the centers of the cells containing $u$ and $v$ respectively and
\begin{eqnarray}
\label{eq:ClocStationaryM}
\Iloc & = & V \times\int_{D\times D}\esp\left[C(\|u-X_{k_0}\|)+C(\|v-X_{k_n}\|)\right]\,f(u)\,f(v) \,du\,dv \nonumber \\
& = & V\;\mu^2\;\kappa^{\textrm{\tiny loc}}\;,
\end{eqnarray}
where $\mu:=\int f$ is the average total number of users and, by stationarity of the point process $M$,
\begin{equation*}
\kappa^{\textrm{\tiny loc}}= \esp\left[C(\|u-X_{k_0}\|)\right]+ \esp\left[C(\|v-X_{k_n}\|)\right]= 2\;\esp\left[C(\|X_0\|)\right]
\end{equation*}
with $X_0$ defined as the center of the cell containing the origin. Note that $\kappa^{\textrm{\tiny loc}}$ denotes the average local cost per secret bit and per pair of users. If $M$ is a Poisson point process with intensity $\alphabb^{-2}$, we further have
\begin{equation*}
\pr(\|X_0\| > t)=\pr(\#\{X_k\;:\;\|X_k\|\leq t\}=0)=\exp(-\pi  t^2\alphabb^{-2})\;,
\end{equation*}
and hence
\begin{equation}
\label{eq:kappalocHomPoissonM}
\fl \;\;\;\; \kappa^{\textrm{\tiny loc}} = 4\pi\alphabb^{-2} \; \int_{\Rset_+} C(t) \; t \; \exp(-\pi t^2\alphabb^{-2}) dt = 4\pi \; \int_{\Rset_+} C(\alphabb u) \; u \; \exp(-\pi u^2) du \; .
\end{equation}

For $\Ibb$, we can write
\begin{equation*}
\Ibb=V\times\int_{D\times D}\esp\left[\sum_{i=1}^n C(\|X_{k_{i}}-X_{k_{i-1}}\|)\right]\,f(u)\,f(v) \,du\,dv\;.
\end{equation*}
Applying~\cite[Theorem~2]{baccelli:aap00} or the results (in particular Theorem~2.41 and Remark~2.4.2) in section~2.4 of \cite{tchoumatchenko:phd99} (as done in Corollaries~2.5.1 and~2.5.2 in~\cite{tchoumatchenko:phd99}), we obtain
\begin{equation*}
\esp\left[\sum_{i=1}^n C(\|X_{k_{i}}-X_{k_{i-1}}\|)\right]= \kappa^{\tiny bb}\, \|u-v\| \, ,
\end{equation*}
where
\begin{equation}
\label{eq:kappaBBHomPoissonM}
\fl \kappa^{\tiny bb} :=   2\alphabb^{-1} \int_{(r,\psi,\phi)\in\mca} C\left(2\alphabb r\sin(\{\psi-\phi\}/2)\right) \,\{\cos(\phi)-\cos(\psi)\}\,r^2\,\rme^{-\pi\,r^2}\,d\psi\,d\phi\,dr \; ,
\end{equation}
and $\mca=\Rset_+\times\{(\psi,\phi):\,0<|\phi|\leq\psi<\pi\}$.
Finally we find that
\begin{equation}
\label{eq:CBBStationaryM}
\Ibb=V\,\kappa^{\textrm{\tiny bb}}\, \delta\;,
\end{equation}
where $\delta$ is the average total distance between two different users defined in equation~(\ref{eq:deltaDef}) and computed in equation~(\ref{eq:deltaVal}), and $\kappa^{\textrm{\tiny bb}}$ denotes the average backbone cost per secret bit and per length unit of the distance separating a pair of users.

From equations~(\ref{eq:totalCost}),~(\ref{eq:ClocStationaryM}) and~(\ref{eq:CBBStationaryM}), and observing that here
the average total number of backbone cells $N^2=(L/\alphabb)^2$, we find
\begin{equation}
\label{eq:totalCostStationary}
\mcc =: \Iloc +\Ibb + \mcc^{\textrm{\tiny node}}=V\times\left[\mu^2\kappa^{\textrm{\tiny loc}}+\delta\kappa^{\textrm{\tiny bb}}\right] + C_{\textrm{\tiny node}}(L/\alphabb)^{2}\ \;,
\end{equation}
where $\mu^2$ and $\delta$ are related to the spatial distribution of the users, and $\kappa^{\textrm{\tiny loc}}$ and $\kappa^{\textrm{\tiny bb}}$ are constants related to the geometry of the backbone and to the routing policy. For users uniformly distributed in a square of side length $L$ with intensity $\alphau^{-2}$, we have $\mu^2\simeq (L/\alphau)^4$ and $\delta\simeq L^5/\alphau^4$.


Using~(\ref{eq:kappalocHomPoissonM}),~(\ref{eq:kappaBBHomPoissonM}),~(\ref{eq:totalCostStationary}) and the above
approximations of $\mu^2$ and $\delta$, we see that the total cost $\mcc$ only depends on $L$, $\alphau$ and $\alphabb$.
Now, for given $\alphau$ and $L$, we take $\alphabb$ so that $\mcc$ is minimized and examine
which term in the right-hand side of~(\ref{eq:totalCostStationary}) dominates the total cost $\mcc$ as $L\to\infty$ in this
context. To this end, we first study each term separately. We let $c$ denote a constant not depending on $L,\alphabb$ in the following
reasoning. Observe that since $C$ is convex and increasing,  $C(\ell)\geq
c\times \ell$. Using this in~(\ref{eq:kappalocHomPoissonM}) and in~(\ref{eq:kappaBBHomPoissonM}),
we get $\Iloc\geq c\,\alphabb L^4$ and $\Ibb\geq c\, L^5$, respectively. Concerning the last
term, we have $\mcc^{\textrm{\tiny node}}\approx c\, L^2/\alphabb^2$. It follows that at fixed $L$, $\Iloc\to\infty$ as
$\alphabb\to\infty$ and $\mcc^{\textrm{\tiny node}}\to\infty$ as $\alphabb\to0$, from which we can deduce that the optimal
$\alphabb$ stays away of 0 and $\infty$. Now, clearly, if $\alphabb$ stays away from 0 and $\infty$, the above bounds show that
$\Ibb$ dominates as $L\to\infty$. Hence, for large $L$, the optimal intensity $\alphabb$ is the one that minimizes $\Ibb$
or, equivalently, $\kappa^{\textrm{\tiny bb}}$. To find this optimal intensity, the following result is useful for an exponential cost $C(\ell)
= \frac{C_{\textrm{\tiny QKD}}}{R_0} e^{\, \ell/\lambda_{\textrm{\tiny QKD}}}$:

\begin{lem}
\label{lem:kappaLocComp}
Define $\kappa^{\textrm{\tiny bb}}$ as in equation~(\ref{eq:kappaBBHomPoissonM}) with $C(\ell) = \frac{C_{\textrm{\tiny QKD}}}{R_0} e^{\, \ell/\lambda_{\textrm{\tiny QKD}}}$. Then the following analytical formula holds
\begin{equation*}
\kappa^{\textrm{\tiny bb}}=C_{\textrm{\tiny QKD}} R_0^{-1} \lambda_{\textrm{\tiny QKD}}^{-1}
\frac4\pi\left[ \rme^{\alphabb^2/(\pi \lambda_{\textrm{\tiny QKD}}^2)}\{1+\erf(\alphabb/(\sqrt{\pi}\lambda_{\textrm{\tiny QKD}}))\}+
\lambda_{\textrm{\tiny QKD}}/\alphabb\right]\;,
\end{equation*}
where
\begin{equation*}
\erf(x)=\frac{2}{\sqrt{\pi}}\int_0^x\rme^{-t^2}\,dt\;.
\end{equation*}
\end{lem}

\noindent \emph{Proof.} Let $s= \lambda_{\textrm{\tiny QKD}}/\alphabb$. We have 
\begin{eqnarray*}
& & \int_{(r,\psi,\phi)\in\mca} \exp\left(2s^{-1} r\sin(\{\psi-\phi\}/2)\right)
\,\{\cos(\phi)-\cos(\psi)\}\,r^2\,\rme^{-\pi\,r^2}\,d\psi\,d\phi\,dr \\
& & = 8\int_{v=0}^{\pi/2}\int_{r=0}^{\infty}\exp(2s^{-1}r\sin(v)-\pi r^2)\,r^2\,\sin(v)\,dv\,dr .
\end{eqnarray*}
Integrating with respect to $r$ yields
\begin{eqnarray*}
\fl \kappa^{\textrm{\tiny bb}} & = C_{\textrm{\tiny QKD}} R_0^{-1} \lambda_{\textrm{\tiny QKD}}^{-1} \\
\fl & \times\left[\frac2\pi+\frac{4s}{\pi} \int_{v=0}^{\pi/2}\sin(v)\{1+2\sin^2(v)/(\pi s^2)\}\exp(\sin^2(v)/(\pi s^2))(1+\erf(\sin(v)/(\sqrt{\pi}s)\,dv\right]\;.
\end{eqnarray*}
Further computations yield
\begin{equation*}
\kappa^{\textrm{\tiny bb}}=C_{\textrm{\tiny QKD}} R_0^{-1} \lambda_{\textrm{\tiny QKD}}^{-1}
\frac4\pi\left[ \rme^{1/(\pi s^2)}\{1+\erf(1/(s\sqrt{\pi}))\}+s\right]\;,
\end{equation*}
which is the desired expression.\\

Using Lemma~\ref{lem:kappaLocComp}, the $\alphabb$ minimizing $\kappa^{\textrm{\tiny bb}}$, denoted as $\alphabb^{\textrm{\tiny opt}}$ below, can easily be calculated using a numerical procedure. We find
\begin{equation}
\alphabb^{\textrm{\tiny opt}} \approx 1.2490 \,  \lambda_{\textrm{\tiny QKD}} \;.
\end{equation}
This result should be compared with the result of equation~(\ref{eq:alphaOptSquareBB}), where the backbone geometry is deterministic and also characterized by the node intensity $1/ \alphabb^2$. The two results show that the choice of the backbone and routing policy does influence the optimal node intensity, albeit in a modest way.


\subsection{From cost optimization results to QKD network planning}
\label{subsec:QKDnetplanning}

\noindent \emph{Matching QKD network topology with QKD links optimum working distance}

\noindent The calculations in sections~\ref{subsec:square} and \ref{subsec:stochasticQBB} point to one common result: it appears that, for large networks, the costs associated with the QKD devices that have to be deployed in backbone nodes to serve the demand are always dominant over the  local costs, associated to the end connections between QKD users and backbone nodes.

Moreover, the optimization of backbone costs indicates that minimum cost will be reached when the typical distance between backbone nodes is of the order of $\lambda_{\textrm{\tiny QKD}}$, the scaling parameter of the curve $R(l)$.

These results lead to the following statements:
\begin{itemize}
\item When a QKD network deployment is planned, is seems optimal to choose the location of network nodes so that QKD links will be operated over distances comparable to the optimal distance $\ell^{\textrm{\tiny opt}}$. As we have seen in our different models, $\ell^{\textrm{\tiny opt}}$ is always lower bounded by a pre-factor times $\lambda_{\textrm{\tiny QKD}}$. Indeed, when the total cost of node equipment can be neglected compared to the cost of QKD devices, as it is the case for large networks, then the optimum distance $\ell^{\textrm{\tiny opt}}$  is indeed comparable to $\lambda_{\textrm{\tiny QKD}}$, which is roughly equal to 20~km. This indicates that current QKD technologies, for which $D_{\textrm{\tiny max}}$ is already significantly larger than 20~km, are well suited for metropolitan operation. On the other hand, the typical distance between amplifiers, in optical wide area networks, is of the order of 80~km. If we wanted to deploy trusted QKD networks with the current generation of QKD devices, the QKD links would have to be operated close to their maximum distance, where the unit of secret bit rate becomes very expensive. Although technically already feasible, the deployment of wide area QKD networks thus remains a challenge. We can however anticipate that this challenge will be overcome within the next years, as new generations of QKD protocols and devices, able to generate keys at higher rates, and with larger maximum distances are already being presented~\cite{stucki:qp08, leverrier:qp08, dixon:qp08}.
\item The results on cost minimization that we have obtained could provide some helpful guidelines for QKD device developers: they may help promoting the idea that what will really matter, in the perspective of real network deployment, will be to focus on the optimization of their systems around typical network-optimum working distances. Optimizing QKD devices in this regime means reducing the cost of a unit bit rate at a \emph{reasonable} distance, where the throughput of the QKD link is not considerably smaller than $R_0$. It will be of course always profitable to design QKD devices that can reach very long distances, but as discussed in~\cite{alleaume:inprep09}, from a system development point of view it can be significantly different to optimize QKD devices to reach the longest possible distance $D_{\textrm{\tiny max}}$, and to optimize them so that the cost of unit of bit rate is as low as possible, around the distance $\ell^{\textrm{\tiny opt}}$ minimizing network costs.\\
\end{itemize}

\noindent \emph{In which regime are backbones useful?}

\noindent We would like now to use our calculation results to analyze in which regime QKD backbones become \emph{economically interesting}, \emph{i.e.} under which conditions it is interesting to introduce some hierarchy and resource mutualization in QKD networks, in order to decrease the total deployment cost.

In the previous sections we have performed cost calculations that can be used to establish some quantitative comparisons between:
\begin{itemize}
\item The cost of a QKD network with no hierarchy as in the generalized linear chain QKD network, whose cost calculations have been performed in section~\ref{subsec:2Dchain}.
\item The cost of a QKD network with one level of hierarchy, which is the case of the square backbone QKD network studied in section~\ref{subsec:square}.
\end{itemize}

Since these two cost calculations have been performed under the same assumptions regarding user distribution and traffic demand, we can use the results given in equations~(\ref{eq:Cchaintotal}) and (\ref{eq:CbbSquare}) to compare the  total network deployment costs, respectively for the generalized linear chain model and for a QKD network with a square backbone (for which we have seen that we could neglect the cost of the local access network).

The condition under which it will be more cost effective to deploy a quantum backbone than to connect all pair of users by one-dimensional chains of QKD links can be described by the following inequality between the respective optimal costs
\begin{eqnarray}
 \fl \mcc_{\textrm{\tiny 2D,chain}}^{\textrm{\tiny opt,chain}} \geq \mcc_{\textrm{\tiny 2D,square}}^{\textrm{\tiny opt,square}} \nonumber \\
 \fl  \Leftrightarrow    \Big( V \, C(\ell^{\textrm{\tiny opt}})/\ell^{\textrm{\tiny opt}}  +  C_{\textrm{\tiny node}}/\ell^{\textrm{\tiny opt}}     \Big)  \gamma \sigma^2 L^5 \, \geq \,   \frac{2}{3}  \, C(\alphabb^{\textrm{\tiny opt}})/\alphabb^{\textrm{\tiny opt}}  \, \sigma^2 L^5 \, V  + \, C_{\textrm{\tiny node}} \, {L^2}/{{\alphabb^{\textrm{\tiny opt}}}^2}   \label{ineq:chainsquare1}
\end{eqnarray}

The above equation is not very convenient to handle because in general $ \alphabb^{\textrm{\tiny opt}} \neq \ell^{\textrm{\tiny opt}} $. However,
\begin{eqnarray}
 \mcc_{\textrm{\tiny 2D,chain}}^{\textrm{\tiny opt,chain}} \geq \mcc_{\textrm{\tiny 2D,square}}^{\textrm{\tiny opt,square}}  \Rightarrow  \mcc_{\textrm{\tiny 2D,chain}}^{\textrm{\tiny opt,square}} \geq \mcc_{\textrm{\tiny 2D,square}}^{\textrm{\tiny opt,square}} \label{implication:chainsquare1}
\end{eqnarray}

Thus, we can derive a necessary condition under which the deployment of a backbone for a QKD network is a better solution than a design that would solely rely on the generalized linear chain of QKD links to transport the traffic:
\begin{eqnarray}
\fl \mcc_{\textrm{\tiny 2D,chain}}^{\textrm{\tiny opt,square}} \geq \mcc_{\textrm{\tiny 2D,square}}^{\textrm{\tiny opt,square}}   \Leftrightarrow   \, C_{\textrm{\tiny node}} \, (  \sigma^2 L^3 \alphabb^{\textrm{\tiny opt}} \,  \gamma - 1 )  \,  \geq   \, C(\alphabb^{\textrm{\tiny opt}}) V \,  \sigma^2 L^3 \alphabb^{\textrm{\tiny opt}} \,  (\frac{2}{3} -   \gamma) \nonumber   \\
 \fl  \Leftrightarrow   \,C_{\textrm{\tiny node}} \, ( \sigma^2/ {\sigma^\ast}^2  - 1)   \,    \geq \,   C(\alphabb^{\textrm{\tiny opt}}) \,  V  \,  \sigma^2/ {\sigma^\ast}^2 \,   ( \frac {2}{3 \gamma} - 1)  \label{implication:chainsquare2}
\end{eqnarray}

with $\sigma^{\ast} =  1/ \sqrt{ L^3 \alphabb^{\textrm{\tiny opt}} \,  \gamma}\; .$\\

Keeping in mind that $\frac {2}{3 \gamma} - 1$ is a positive number, we can use the last inequality to make the following observations:
\begin{itemize}
\item First, it appears that, if the user density $\sigma$ is smaller than $\sigma^{\ast}$, which we can qualify as a \emph{critical user density}, then equation~(\ref{implication:chainsquare2}) can never be verified. This means that below $\sigma^\ast$ it will never be interesting to deploy a backbone. This result has a clear interpretation: backbone infrastructures can only be interesting in the case where sharing resources offers a cost reduction. And the incentive to share a backbone infrastructure can only exist if there are enough users. The minimum total number of users required to have a cost incentive towards backbone deployment is $\sigma^{\ast}\, L^2 = \sqrt{ L / (\gamma \alphabb^{\textrm{\tiny opt}})} $.
\item In case $\sigma$ is larger than the critical user density $\sigma^{\ast}$, we enter a regime where there will be an incentive to deploy a quantum backbone essentially if the cost of a node $C_{\textrm{\tiny node}}$ dominates over the cost of QKD link equipment to be deployed, which scales as $C(\alpha_{bb}^{\textrm{\tiny opt}}) V$. This also has a clear interpretation: if we take the extreme case where building a node (and installing node equipment inside it) is zero, we can foresee that there will be no incentive to build a backbone: it will always be cheaper to deploy direct chains between each pair of users. The motivation to build a backbone arises when efforts associated to opening a QKD node are important. This will of course be the case if QKD node equipment is expensive, as we can see from equation~(\ref{implication:chainsquare2}), but it is also intuitive that, in case significant efforts are required to build new QKD nodes, mutualization of nodes through a backbone structure will be a cost effective solution.
\end{itemize}

\section{Conclusion and Perspectives}
\label{sec:perspectives}

In this paper, we performed a topological analysis of quantum key distribution networks with trusted repeater nodes. In particular, under specific assumptions on the user and node distributions as well as the call traffic and routing in such networks, we derived cost functions for different network architectures. We first considered a linear chain network as a basic model that served the purpose of illustrating the main techniques and ideas that we used, and then moved on to more advanced network configurations that were in some cases enhanced with a backbone structure. Using cost minimization arguments, we obtained results on the optimal working points of QKD links and spatial distribution of QKD nodes, and examined the importance of introducing hierarchy into QKD networks.

Our results indicate that, in the context of QKD networks, it is more cost-effective and therefore advantageous to operate individual QKD links at their optimal working point, which is in general significantly shorter than the maximum span of such links. This conclusion motivates the research of new experimental compromises in practical QKD systems, and can be illustrated by considering examples of such systems where the characteristics of either a hardware component (for example a single-photon detector) or a software algorithm (for example a reconciliation code) can be experimentally manipulated as a function of distance~\cite{alleaume:inprep09}.

In general, it is clear that, as the realization of more and more advanced QKD networks approaches the realm of actual deployment, it becomes necessary to orient the research on QKD devices and links towards cost-related directions, and extend the techniques we have presented here to more sophisticated network technologies and architectures.

\ack
We acknowldge financial support from the Integrated European Project SECOQC (Grant No. IST-2002-506813). R. A. and E. D. acknowledge financial support from the French National Research Agency Projects PROSPIQ (ANR-06-NANO-041-05) and SEQURE (ANR-07-SESU-011-01). N. L. acknowledges support from the NSERC Innovation Platform QuantumWorks, a NSERC Discovery Grant, and the Ontario Centers of Excellence.

\end{document}